\pdfoutput=1
\documentclass[11pt]{article}
\usepackage[margin=1in]{geometry}
\usepackage[T1]{fontenc}
\usepackage{newtxtext}
\usepackage[varvw]{newtxmath}
\usepackage{cite}
\usepackage{hyperref}

\usepackage{graphicx}
\usepackage{booktabs}
\usepackage{cleveref}
\newcommand{\DS}[1]{\texttt{#1}}
\hypersetup{colorlinks,allcolors=blue,
  pdfauthor={Mihir Agarwal, Sabyasachi Basu, Viraj Agrawal},
  pdftitle={Simple Linear Corrections for Compression in Late Interaction}}

\title{Simple Linear Corrections for Compression in Late Interaction}

\author{Mihir Agarwal\thanks{Equal contribution.} \qquad Sabyasachi Basu\footnotemark[1] \qquad Viraj Agrawal \\
  Microsoft Research India \\
  \small\texttt{\{t-miagarwal, sabyasachi.basu, t-viragrawal\}@microsoft.com}
}

\date{}

\begin{document}

\maketitle

\begin{abstract}
Late-interaction (multi-vector) models represent documents using a set of token-level vectors, providing a middle ground between cross-encoders and dense encoders in terms of quality and cost of embedding. However, compared to the latter, the indices are costly to store and score. Pooling and pruning based compression techniques are deployed widely but deteriorate in quality noticeably with aggressive compression. We investigate whether a lightweight change to the scoring function can recover retrieval quality
after document-vector compression. For each combination of encoder, compressor, and vector budget, we learn a global linear correction by distilling the uncompressed model's score distributions over candidate documents. The encoder and compressor remain frozen, and the correction can be applied to each query at search time or once, offline, to the stored vectors.

We evaluate the method with ColBERTv2 and Jina v4 on \DS{MS MARCO} and \DS{BEIR}, and with ColPali2 on the \DS{ViDoRe v3} visual document retrieval benchmark. Experiments cover four training-free compression methods, multiple vector budgets, in-domain adaptation, and transfer from \DS{MS MARCO} to \DS{BEIR}. In the in-domain setting, the correction improves over baseline compression in 154 of 160 \DS{BEIR} settings, while
transfer from \DS{MS MARCO} is more variable across models and datasets. We also find that matrices trained with synthetic queries can improve over baseline compression on \DS{ViDoRe v3}. These results show that a small learned correction can recover some of the quality lost to compression without retraining the encoder or recompressing the documents.
\end{abstract}

\section{Introduction}

Late-interaction models, introduced by Khattab and Zaharia~\cite{khattab2020colbert}, represent queries and documents as sets of token-level vectors. This representation offers several advantages, including greater robustness to catastrophic forgetting~\cite{SAGKS26}, improved generalization in out-of-domain settings~\cite{thakur2021beir,santhanam2022colbertv2}, and substantially stronger performance on multimodal retrieval~\cite{faysse2024colpali}. More recently, theoretical and empirical work has also exposed fundamental limitations of single-vector representations relative to their multi-vector counterparts~\cite{weller2025,SAGKS26,j26,agarwal2026retrieval}. However, storing and scoring many vectors for every document is costly, particularly when a multi-vector model is used to rerank or search over a large candidate set.

Multi-vector models also play a crucial role in visual document retrieval, where textual queries are used to search over images and screenshots~\cite{faysse2024colpali, moreira2026nemotron, kolavi2025m3dr, gunther2025jina}. In this case, models represent documents using an even larger number of vectors (usually around 1000 per document). This further motivates the question of reducing the representation size (in terms of number of vectors) of multi-vector retrieval models.

Storage can be substantially reduced through aggressive quantization~\cite{santhanam2022colbertv2,jegou2011product}, but scoring remains expensive: the MaxSim (Chamfer) similarity requires computing pairwise dot products between query and document tokens and taking a maximum over document tokens for each query token. Systems such as PLAID~\cite{santhanam2022plaid} mitigate this cost by pruning pairwise interactions. A common way to reduce these costs is to represent each document with fewer vectors. Compression reduces index size and scoring work, but can alter the similarities used to rank documents. This trade-off is especially important at small vector budgets.

Many of these methods involve training the representation to allow meaningful, succinct summaries~\cite{macavaney2025efficient, xiao2025metaembed, qin2026multi}. An alternative group of common techniques involves pooling or clustering token embeddings after training, typically by some pooling or pruning technique~\cite{lassance2021study, acquavia2023static, he2025token, clavie2024reducing, yan2025docpruner, yan2026sculpting, yan2026visual}. $k$-means or Ward's clustering~\cite{kmeans1982,ward1963} are reported to be among the most effective~\cite{clavie2024reducing, jha2026brief}. More recently, EigenLI~\cite{s2026eigenlispectralapproximationslate} exploits the low-rank structure of token embeddings through a PCA-like reduction.
Such compression is especially useful in multi-stage retrieval, where a multi-vector model reranks candidates first retrieved by a single-vector or learned sparse retriever~\cite{MNRV26}.

However, compression costs quality, especially at small budgets. With ColBERTv2 on \DS{MS MARCO}, Ward compression to 32 vectors per document costs only about one point of Recall@10, but compression to 4 vectors costs over 20 points, about a third of the original score. We therefore ask whether a lightweight correction to the scoring function can recover some of this loss without changing the encoder or the compressor.

\subsection{Our Contributions}

We propose a simple post-hoc approach for recovering retrieval quality under aggressive compression. We learn one global linear transformation for each encoder, compressor and vector budget, keeping the encoder frozen, by distilling the uncompressed model's MaxSim scores on a few thousand training queries. On \DS{MS MARCO} at $k=4$, the correction improves capped Recall@10 over baseline compression by up to 22\% for ColBERTv2 and 32\% for Jina v4, depending on the compression method. For in-domain \DS{BEIR} adaptation, the relative gain is positive in 79 of 80 settings for ColBERTv2 and 75 of 80 for Jina v4, with median gains of about 10\% and 9\%. Transfer from \DS{MS MARCO} is less reliable: it improves most \DS{BEIR} settings for Jina v4 by dramatic margins, routinely exceeding 20\% in relative gains and crossing 40\% in some cases. However, such improvements in the out-of-domain setting are observed for only a little over half of the setups using ColBERTv2. On \DS{ViDoRe v3}, in-domain correction improves dataset-averaged Recall@10 for every compressor and budget, by 1.0--10.2\%.

The correction can improve all four compression methods we test, by amounts that depend on the method, and needs no recompression. For the latency study, we apply $M$ once, offline, to the stored vectors, so reranking does the same work as baseline compression. Alternatively, $M$ can be applied to each query token at search time, costing one $128\times128$ matrix--vector product per token while leaving the stored vectors unchanged. The loss uses only the uncompressed model's MaxSim scores as targets. When relevance judgments (qrels) are available, we use them to include the judged positives in each training panel; without them, as with our synthetic queries, the panel holds the source document and the teacher's top-ranked documents. On \DS{FiQA} and \DS{NFCorpus}, LLM-generated queries recover 60--93\% of the human-query gain for ColBERTv2 and about the same gain for Jina v4.

\section{Related work}
\label{sec:related-work}

ColBERT~\cite{khattab2020colbert} represents a query and a document by one
vector per token and scores them with MaxSim: each query vector is matched to its
most similar document vector, and these similarities are summed. Late-interaction
models are strong out of domain~\cite{thakur2021beir,santhanam2022colbertv2}.
Recent work also shows the limits of single-vector
embeddings~\cite{weller2025,SAGKS26} and proves that multi-vector representations
are more expressive than single vectors of the same
size~\cite{KIZM26,j26,jayaram2026near,agarwal2026retrieval}. The same idea is used
for visual documents, where each image patch gets a
vector~\cite{faysse2024colpali,moreira2026nemotron}, and in multimodal embedding
models with multi-vector outputs~\cite{gunther2025jina}. The main cost is the
index, which stores one vector per token. Several systems make storing and
searching these vectors cheaper. ColBERTv2~\cite{santhanam2022colbertv2} stores
each vector as its nearest centroid plus a quantized residual.
PLAID~\cite{santhanam2022plaid} uses these centroids to discard most candidate
documents before exact scoring.
MUVERA~\cite{dhulipala2024muvera} turns each set of vectors
into a single fixed-dimensional vector, so that standard single-vector search can
find candidates. These systems change how the stored vectors are searched, not
how many are stored, and are complementary to the methods below.

Pruning keeps only some of a document's token vectors, chosen by heuristics such
as position, IDF or attention scores~\cite{lassance2021study,acquavia2023static}
or by a learned classifier~\cite{he2025token}. Pooling merges vectors instead of
dropping them. Clavi\'e et al.~\cite{clavie2024reducing} cluster each document's
vectors with $k$-means~\cite{kmeans1982} or Ward's hierarchical
clustering~\cite{ward1963} and keep the mean of each cluster; Ward works best and
halves the number of vectors with almost no loss. A recent comparison of
training-free methods finds that merging works better than
pruning~\cite{jha2026brief}. EigenLI~\cite{s2026eigenlispectralapproximationslate}
keeps a subspace instead of a set of vectors: it stores the top eigenvectors of
each document's token second-moment matrix and scores a query token by the
squared length of its projection onto them, an idea with roots in
quantum-inspired
IR~\cite{piwowarski2009quantum,zuccon2009semantic,piwowarski2010filtering,caputo2011query}.
For visual documents, DocPruner~\cite{yan2025docpruner} adaptively prunes patch
vectors on each page, Yan et al.~\cite{yan2026sculpting} prune and then merge,
and ColChunk~\cite{yan2026visual} clusters patch vectors hierarchically using
their 2D positions. Other methods change the encoder.
ConstBERT~\cite{macavaney2025efficient} learns a projection from the token
vectors to a fixed number of vectors; MetaEmbed~\cite{xiao2025metaembed} appends
learnable meta tokens whose outputs form the document representation; and Qin
et al.~\cite{qin2026multi} compare memory tokens, hierarchical pooling and
attention-guided clustering. Learn to Pool~\cite{josef2026learn} and Chaffin et
al.~\cite{chaffin2026hpoolregularization} instead fine-tune or regularize the
encoder so that its vectors pool better. Finally, each vector can itself be made
smaller, by quantization~\cite{pan2023surveyvectordatabasemanagement} or dimension
reduction~\cite{siciliano2024static}.

Metric learning replaces a fixed similarity with a
learned one defined by a positive semidefinite (PSD) matrix $M$, for example the
Mahalanobis distance $(x-y)^{\top}M(x-y)$. Information-Theoretic Metric Learning
(ITML)~\cite{itml2007} learns $M$ while keeping it close to a prior such as the
identity. Large Margin Nearest Neighbor (LMNN)~\cite{weinberger2009} learns $M$
so that each point's nearest neighbors share its class, while points from other
classes are kept a margin away. Neighborhood Components Analysis
(NCA)~\cite{goldberger2004} learns a linear map $A$, giving $M=A^{\top}A$, that
improves nearest-neighbor classification.

\section{Late Interaction and Document Compression}
\label{sec:compression}
A late-interaction model encodes a query into token embeddings
$Q=\{q_1,\dots,q_m\}$ and a document into $D=\{x_1,\dots,x_n\}$, with
$q_j,x_i\in\mathbb{R}^d$, and scores the pair with the sum-MaxSim (Chamfer)
similarity
\begin{equation}
  S(Q,D)=\sum_{j=1}^{m}\max_{1\le i\le n} q_j^{\top}x_i .
  \label{eq:maxsim}
\end{equation}
A compressor replaces the $n$ vectors of each document with $k\ll n$ vectors
$C=\{c_1,\dots,c_k\}$. It does this offline and independently per document,
without access to queries, so storage and scoring cost grow with $k$ rather than
$n$. We study four training-free compressors. For $n\le k$, Ward and
$k$-means retain the $\ell_2$-normalized original vectors, while FPS retains
the original vectors; EigenLI still constructs a spectral basis.

\noindent\textbf{Ward.}
Ward's method~\cite{ward1963} is agglomerative hierarchical clustering. It starts
with every token as its own cluster and repeatedly merges the two clusters whose
union least increases the total within-cluster sum of squared Euclidean
distances, stopping when $k$ clusters remain. Each cluster is represented by its
$\ell_2$-normalized centroid.

\noindent\textbf{\boldmath $k$-means.}
$k$-means~\cite{kmeans1982} also minimizes the within-cluster sum of squares,
but it reassigns tokens directly. Lloyd's algorithm alternates between assigning
each token to its nearest centroid and recomputing the centroids. We initialize
with $k$-means++~\cite{arthur2007} from a fixed seed, use a single restart and stop after at most
300 iterations or at convergence. As with Ward, each document is represented by
its $k$ $\ell_2$-normalized centroids.

\noindent\textbf{Farthest point sampling (FPS).}
FPS~\cite{gonzalez1985} selects tokens rather than averaging them. It starts
from a randomly chosen token and repeatedly adds the token with the lowest
maximum cosine similarity to the tokens already selected. The result is a
diverse subset that covers the document; this greedy rule is a classical
2-approximation to the $k$-center problem. The kept vectors are original
embeddings, so a query token that matches a kept token scores exactly as it
would without compression.

\noindent\textbf{EigenLI.}
EigenLI~\cite{s2026eigenlispectralapproximationslate} uses the low-rank
structure of a document's token embeddings. It keeps the top-$k$ eigenvectors
$u_1,\dots,u_k$ of the uncentered second-moment matrix $\sum_i x_i x_i^{\top}$,
which span the $k$-dimensional subspace that best approximates the tokens in the
least-squares sense. Directions with numerically zero eigenvalues are dropped,
so a document keeps at most $\min(n,d)$ vectors. Since eigenvectors are defined
only up to sign, EigenLI replaces the maximum in \Cref{eq:maxsim} with the
projection energy
\begin{equation}
  S_{\mathrm{E}}(Q,U)=\sum_{j=1}^{m}\sum_{c=1}^{k}(q_j^{\top}u_c)^2 ,
  \label{eq:eigenli}
\end{equation}
which rewards query tokens that lie in the document's dominant subspace. When
$k=d$ the basis is complete, so $S_{\mathrm{E}}$ reduces to
$\sum_j\|q_j\|^2$ and no longer depends on the document.

\subsection{The Correction}
\label{sec:correction}
Compression changes what a document vector means to MaxSim. Consider a query token $q$ and a cluster
$G$ of document tokens that Ward or $k$-means replaces with its centroid $c$.
Without compression, MaxSim scores $q$ by its best match in the cluster,
$\max_{i\in G} q^{\top}x_i$. After compression, it uses $q^{\top}c$, which (up
to normalization) reflects the \emph{average} match over the cluster rather than
the \emph{best} one. How much the two differ depends on how spread out the
tokens in $G$ are along the direction of $q$. Our intuition is that, because all
documents come from the same encoder, these errors are similar across documents.
A single matrix that reweights directions of the embedding space could then
correct part of them. We therefore learn a
single matrix $A\in\mathbb{R}^{d\times d}$ for each encoder, compressor and
budget $k$, and apply it to both query tokens and compressed vectors:
\begin{equation}
  S_A(Q,C)=\sum_{j=1}^{m}\max_{c\in C}\,(Aq_j)^{\top}(Ac)
          =\sum_{j=1}^{m}\max_{c\in C}\,q_j^{\top}Mc,
  \qquad M=A^{\top}A\succeq 0 .
  \label{eq:score}
\end{equation}
The matrix $M=A^{\top}A\succeq0$ defines a PSD bilinear similarity, $q^{\top}Mc$; $A=I$ recovers the baseline compressed score. For EigenLI, we apply the same correction to its projection score, replacing each $q_j^{\top}u_c$ with $q_j^{\top}Mu_c$. Although the centroid argument motivates the correction, \Cref{eq:score} assumes nothing about the compressor and applies unchanged to FPS and EigenLI.  The encoder and compressor remain frozen, and $A$ introduces only $d^2$ trainable parameters (16,384 when $d=128$). Since $M$ is symmetric, $q_j^{\top}Mc=(Mq_j)^{\top}c=q_j^{\top}(Mc)$, so $M$ can be applied either to each query token or once to the stored vectors. We note that the factorization $M=A^{\top}A$ is also used in Neighborhood Components Analysis (NCA) to parameterize a Mahalanobis distance~\cite{goldberger2004}. Unlike NCA, which learns a distance for supervised nearest-neighbor classification, we use the induced similarity $q^{\top}Mc$ within the MaxSim framework and train it to match the uncompressed retriever’s score distributions.

Our learned transformation reweights directions in the embedding space and may correct some of this mismatch when the distortions are systematic across the training examples. It does not reconstruct the discarded token-level information, nor is there a guarantee that a single global transformation will recover the original MaxSim scores for every query and document. Rather, it is learned to make the compressed model’s scores better match the uncompressed model’s scores over the training candidate panels. Empirically, our results demonstrate that the technique is effective.

\subsection{Distillation Objective}
\label{sec:objective}
We fit $A$ by distilling the uncompressed model. For each training query $Q$ we
take a panel $\mathcal{P}_Q$ of candidate documents (\Cref{sec:setup}). On this
panel we compare teacher scores $S(Q,D)$, computed on the full token sets, with
student scores $S_A(Q,C_D)$, computed on the compressed vectors (using corrected
projection energy for EigenLI), and minimize
\begin{equation}
  \mathcal{L}(A)=\frac{1}{|\mathcal{B}|}\sum_{Q\in\mathcal{B}}
  \mathrm{KL}\big(p_Q\,\big\|\,\hat p_Q\big)+\lambda\|A-I\|_F^2,
  \label{eq:loss}
\end{equation}
where $p_Q$ and $\hat p_Q$ are the softmax distributions of the teacher and
student scores over $\mathcal{P}_Q$ (temperature $T=1$), $\mathcal{B}$ is a
minibatch of queries, and $A$ is initialized at $I$. Only teacher scores enter
the loss. Relevance labels, when they exist, decide only which documents form a
panel, so the method also works with synthetic queries and no labels. The
regularizer keeps the corrected score close to the encoder's original geometry.

\section{Experimental Setup}
\label{sec:setup}

\begin{figure}[!htb]
    \centering
    \includegraphics[width=0.8\linewidth,clip]{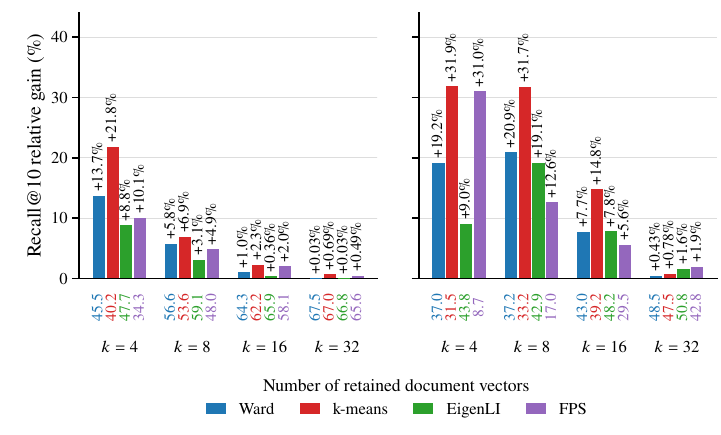}
    \caption{Recall@10 relative gain from $A$ over baseline compression on
    \DS{MS MARCO}: ColBERTv2 (left) and Jina v4 (right). Bars show means.
    Across plotted settings, Recall@10 sample standard deviations (SDs)
    range up to 0.20 percentage points. Values below the bars are baseline
    Recall@10 (\%). Uncompressed Recall@10: 68.84\% for
    ColBERTv2 and 49.08\% for Jina v4.}
    \label{fig:MSMARCO-gains}
\end{figure}

\subsection{Models}
We use three late-interaction encoders. All produce 128-dimensional vectors,
and all are kept frozen. ColBERTv2~\cite{santhanam2022colbertv2} is a BERT-based
text retriever; we use the v2.0 checkpoint and pad queries to 32 tokens.
Jina v4~\cite{gunther2025jina} is a multimodal embedding model built
on a 3B vision--language model; we use its multi-vector output for text
retrieval. ColPali2 (\texttt{colpali2-3b-pt-448}) follows
ColPali~\cite{faysse2024colpali}. It uses a PaliGemma-based vision--language
model to embed each page image as one vector per image patch, and each query as
one vector per token. ColBERTv2 runs in FP32 and the other two run in BF16; all
vectors are stored in FP32.

\subsection{Datasets and evaluation}
For text, we use the \DS{MS MARCO} passage collection~\cite{nguyen2016msmarco} (8.8M
passages) and evaluate on its 6,980 dev queries. From \DS{BEIR}~\cite{thakur2021beir},
we use the five datasets that have a training split: \DS{NFCorpus}, \DS{HotpotQA}, \DS{FiQA},
\DS{FEVER} and \DS{SciFact}. For visual documents, we use ColPali2 on eight
\DS{ViDoRe v3} datasets~\cite{loison2026vidorev3}, comprising 19,252 pages and
2,419 query families with six language versions each. We fix a seed-0 30/70
query-family split within each dataset, keeping translations together: 4,332
training and 10,182 held-out query versions. Both portions use the same complete
corpus of their own dataset, not a union of all eight corpora. This is a custom
in-domain adaptation setting, not the official zero-shot protocol. We equally
average Recall@10 across the eight datasets. Except for the reranking latency
study in~\Cref{sec:latency-setup}, we score each query against every
document in its corpus by brute force, without candidate generation or
approximate search, so differences come only from the scoring function. We
report capped Recall@10, $100\,|\mathrm{Top}_{10}(q)\cap R_q|/\min(10,|R_q|)$,
averaged over queries with at least one relevant document.

\begin{figure}[!htb]
    \centering
    \includegraphics[width=\linewidth,clip]{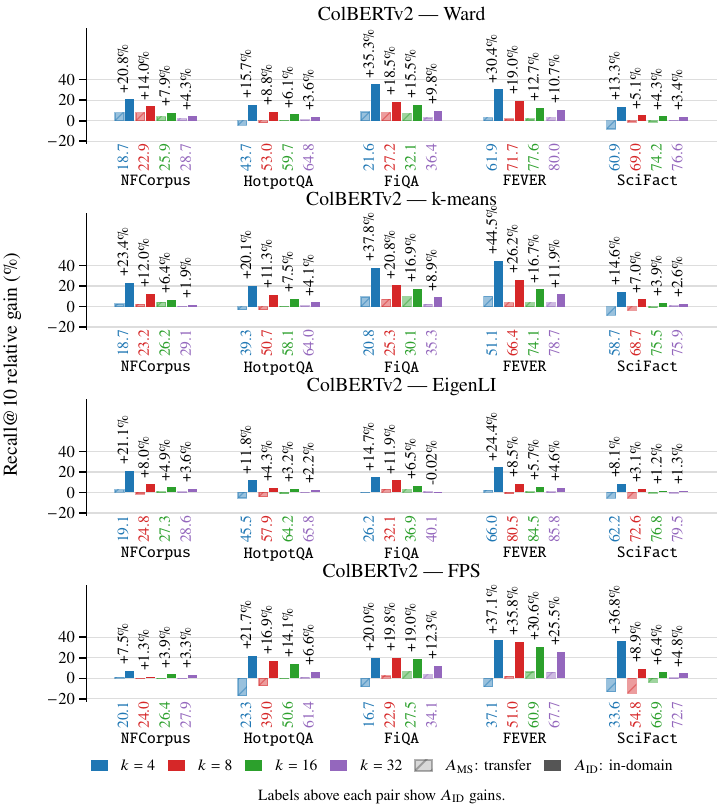}
    \caption{Recall@10 relative gains over baseline compression for \DS{MS MARCO} transfer
    ($A_{\mathrm{MS}}$) and in-domain correction
    ($A_{\mathrm{ID}}$) for ColBERTv2 for different compression schemes.
    Bars show means; across both parts, Recall@10 sample SDs range up to 0.57 percentage points.
    Values below the bars are baseline Recall@10 (\%).
    Uncompressed references (\%): \DS{NFCorpus} 31.58;
    \DS{HotpotQA} 69.12; \DS{FiQA} 44.26; \DS{FEVER} 90.94; \DS{SciFact} 80.46.
    Jina v4 continues on the next page.}
    \label{fig:OOD-gains}
\end{figure}

\begin{figure}[!htb]
    \centering
    \includegraphics[width=\linewidth,clip]{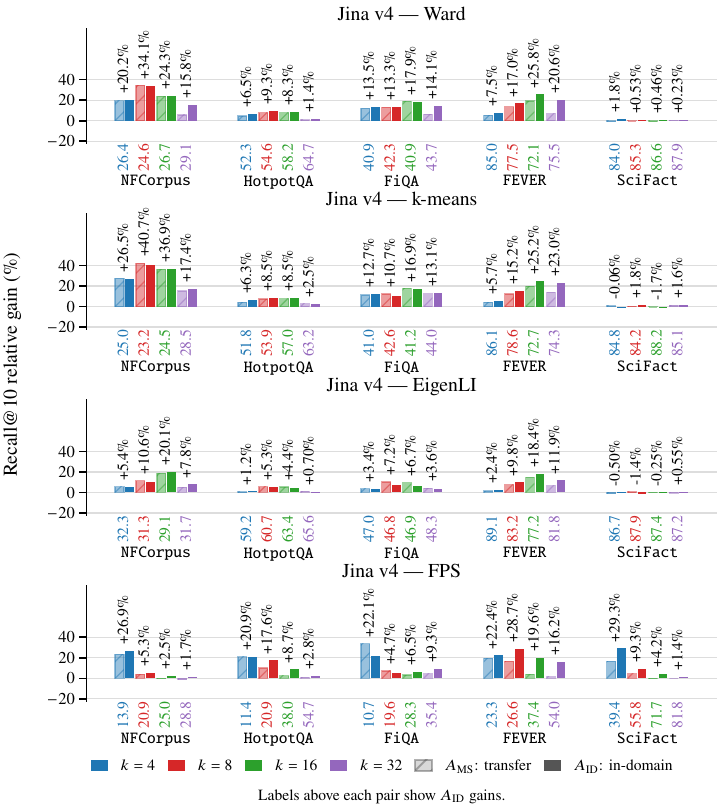}
    \par\vspace{\abovecaptionskip}
    {\small\raggedright
    \textbf{Figure~\ref*{fig:OOD-gains} (continued).}
    Recall@10 relative gains over baseline compression for \DS{MS MARCO} transfer
    ($A_{\mathrm{MS}}$) and in-domain correction ($A_{\mathrm{ID}}$) for Jina v4 for different compression schemes.
    Bars show means;
    across both parts, Recall@10 sample SDs range up to 0.57 percentage points.
    Values below the bars are baseline Recall@10 (\%). Uncompressed references (\%):
    \DS{NFCorpus} 34.55; \DS{HotpotQA} 67.25; \DS{FiQA} 49.13;
    \DS{FEVER} 92.88; \DS{SciFact} 87.82.\par}
\end{figure}

\subsection{Training}
We train a separate matrix $A$ for each encoder, compressor and $k$, with
$k\in\{4,8,16,32\}$. For
text, we train two matrices in each case. The first is trained on 5,000 \DS{MS MARCO}
training queries and applied to \DS{BEIR} unchanged, to test out-of-domain
transfer. Each query gets a 64-document panel with its positives and hard
negatives from the ColBERTv2 training data~\cite{santhanam2022colbertv2}, and a
different panel is used in each epoch; for Jina v4, the same panels are scored by
Jina v4. The second matrix is trained on up to 5,000 queries from the \DS{BEIR}
dataset's own training split, to measure in-domain gains. For ColPali2, we
pool the 4,332 training query versions across the eight \DS{ViDoRe v3} datasets,
weight each version equally, and learn one shared matrix per compressor and
budget. Held-out queries and judgments are excluded from mining and training.
For \DS{BEIR} and ColPali2, a panel contains all positives of the query plus the
63 nonpositive documents that the uncompressed model ranks highest in that
query's own corpus. Every matrix starts at the
identity and is trained with Adam~\cite{kingma2015adam} (learning rate
$10^{-3}$, batches of 64 queries, no weight decay) for five epochs, with
$\lambda=0.01$ and $T=1$. We use the matrix from the last epoch, without early
stopping. We train with five seeds (0--4), which change only the order of
training queries, and report the mean and SD.

\begin{figure}[!htb]
\centering
\includegraphics[width=0.8\linewidth,clip]{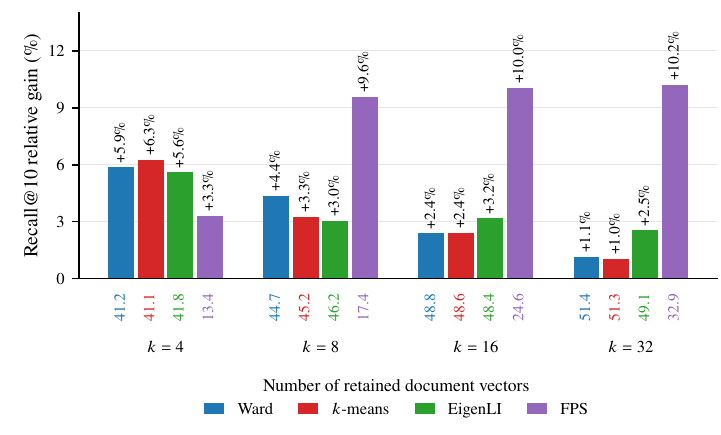}
\caption{ColPali2 Recall@10 relative gains over baseline compression on
  \DS{ViDoRe v3}, with a 30/70 train/test query-family split. Recall@10
  sample SDs range up to 0.10 percentage points. Values below the bars are baseline Recall@10 (\%);
  Full MaxSim Recall@10: 55.12\%.}
\label{fig:multimodal-training}
\end{figure}

\subsection{Synthetic queries}
We also train matrices without any human queries. GPT-6 Luna, with high
reasoning effort, writes one search question for each sampled document of the
target corpus: 5,000 for \DS{FiQA}, 2,590 for \DS{NFCorpus}, and 500 for each of the eight
\DS{ViDoRe v3} subsets, pooled into 4,000. For \DS{NFCorpus}, we ask for the same mix of
question forms and lengths as the official training queries, without showing
the model any human query. For \DS{ViDoRe v3}, the model reads the page image, and a
second pass with the same model rejects questions that the page does not
support. Each question is paired with its source document and the 63 documents
that the uncompressed model ranks highest, so no relevance labels are used.
These matrices use Ward compression and a single seed. For \DS{ViDoRe v3}, we
compare with the matrix trained on all 118,195 \DS{ColPali} training queries,
that is, human queries from a different corpus, and evaluate both on all 14,514
query versions with $k\in\{4,8,16,32,64,128\}$. This is not the in-domain
matrix of \Cref{fig:multimodal-training}, which is evaluated on held-out
queries only.

\begin{figure}[htb]
\centering
\includegraphics[width=0.9\linewidth,clip]{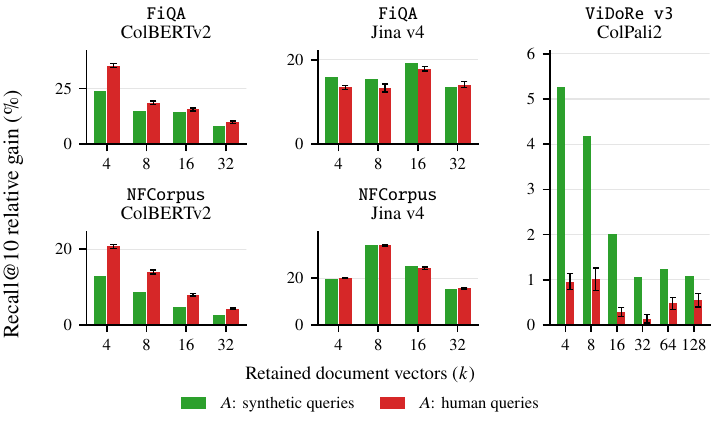}
\caption{Recall@10 relative gains over baseline Ward for synthetic-query
  (left bars) and human-query (right bars) training. Left: text, with human
  queries from the same dataset. Right: ColPali2 on all \DS{ViDoRe v3}
  queries, equally averaging eight subsets; here the human-query matrix is trained on
  \DS{ColPali} queries from a different corpus. Human-query bars show means
  and sample SDs.}
\label{fig:synthetic-queries}
\end{figure}

\subsection{Reranking latency}
\label{sec:latency-setup}
We also measure the cost of reranking with compressed vectors, following the
shortlist setting of Martinico et al.~\cite{MNRV26}. For each of the 6,980
\DS{MS MARCO} dev queries, we rerank the top-50 SPLADE~\cite{splade2022} candidates released by
the authors with ColBERTv2, using full MaxSim, baseline Ward, or Ward corrected
by $A$, with $k\in\{4,8,16,32\}$ and the \DS{MS MARCO} matrices from above.
For corrected Ward, we apply $M$ once, offline, to the compressed document
vectors and cache this transformed copy, so no per-query work for $M$ remains.
All variants use the same FP32 PyTorch scorer on one pinned CPU core, with one
thread and one query at a time, without pruning or early exit. The timer covers
gathering candidate vectors from memory, padding, MaxSim and sorting the top 10;
it excludes encoding, first-stage retrieval and the offline transformation.
We report mean latency over three timed passes after one warm-up pass, and
Recall@10 of the reranked list, counting relevant documents absent from the
shortlist as misses. These are warm-memory reranking times, not end-to-end latency.

\section{Results}
\label{sec:results}
Relative gain is defined as the relative change over baseline compression at the same $k$,
$100(R_A-R_{\mathrm{baseline}})/R_{\mathrm{baseline}}$, where $R_A$ is corrected
Recall@10, averaged over five seeds except for the single-seed synthetic runs.
Absolute differences are reported in percentage points. We do not run significance tests.

\subsection{Text}

\paragraph{\DS{MS MARCO}: }\Cref{fig:MSMARCO-gains} shows that the correction helps most when compression
is most aggressive. At $k=4$, it improves Recall@10 by 8.8\% (EigenLI) to
21.8\% ($k$-means) for ColBERTv2, and by 9.0\% to 31.9\% for Jina v4. For
ColBERTv2, this closes 10--31\% of the gap to the uncompressed model; with
Ward, for example, Recall@10 rises from 45.53\% to 51.75\%, against 68.84\%
without compression. The gains shrink as $k$ grows and stay below 2\% at
$k=32$. The mean improves in all 32 settings, although some changes at $k=32$
are only 0.02 percentage points. Jina v4 gains more than ColBERTv2 at middle budgets: with
Ward at $k=8$, 20.9\% against 5.8\%. With Jina v4, corrected EigenLI at
$k\ge 8$ even scores above the uncompressed model (51.2--52.0\% against
49.08\%), as does baseline EigenLI at $k=32$ (50.75\%).

\paragraph{\DS{BEIR}: }
\Cref{fig:OOD-gains} compares, on each \DS{BEIR} dataset, a matrix trained on
that dataset's training queries (in-domain) with the matrix trained on
\DS{MS MARCO} and applied unchanged (transfer). The in-domain matrix improves
over baseline compression in 79 of 80 settings (five datasets, four compressors
and four values of $k$) for ColBERTv2 and in 75 of 80 for Jina v4, with median
gains of 10.2\% and 8.6\%. The single ColBERTv2 exception is EigenLI at $k=32$
on \DS{FiQA} (40.12\% against 40.13\%). All five Jina v4 exceptions are on
\DS{SciFact}, where baseline $k$-means and EigenLI are already close to the
uncompressed model (84.2--88.2\% against 87.82\%). The largest gains come
at $k=4$, where the median gain for ColBERTv2 is 20.9\% and the maximum is
44.5\% ($k$-means on \DS{FEVER}, from 51.14\% to 73.91\%). With Jina v4 on
\DS{FEVER}, baseline Ward gets worse as $k$ grows, from 84.95\% at $k=4$ to
72.12\% at $k=16$; with the in-domain matrix, it stays between 90.7\% and
91.3\% at every $k$. Even so, corrected compression exceeds the uncompressed
model in only 9 of the 160 in-domain settings.

Transfer is weaker and less consistent. It improves over baseline compression in
49 of 80 settings for ColBERTv2 (median gain 0.5\%) and in 70 of 80 for Jina v4
(median 6.0\%). For ColBERTv2, it mostly hurts on \DS{HotpotQA} and
\DS{SciFact}, where it improves only 5 and 3 of 16 settings; the largest drop is
17.1\% (FPS on \DS{HotpotQA} at $k=4$). The in-domain matrix beats transfer in
79 of 80 ColBERTv2 settings but only in 56 of 80 Jina v4 settings, so Jina v4
transfers much better from \DS{MS MARCO}.

\subsection{Visual document retrieval}
\label{sec:colpali2-results}
\Cref{fig:multimodal-training} shows gains for all 16 compressor--budget
settings after pooled in-domain training. Dataset-averaged Recall@10 improves
by 1.0--10.2\% relative to baseline compression, or 0.44--3.35 percentage points.
At $k=4$, Ward, $k$-means and EigenLI gain 5.6--6.3\% relatively. Improvements
are not universal within individual datasets: 106 of 128 dataset--compressor--budget
settings improve. FPS has the largest relative gains at higher budgets, but its
absolute Recall@10 remains lower: at $k=32$, corrected FPS reaches 36.23\%,
compared with 51.99\% for corrected Ward. All compressed settings remain below
the uncompressed reference of 55.12\%.

\subsection{Synthetic queries}
\label{sec:synthetic-results}
\Cref{fig:synthetic-queries} compares matrices trained on synthetic queries with
matrices trained on human queries. On text, synthetic queries recover most of
the gain without any human query or relevance label. For ColBERTv2, they give
60--93\% of the human-query gain on \DS{FiQA} and \DS{NFCorpus}; on \DS{FiQA}
at $k=4$, for example, Recall@10 rises from 21.65\% to 26.86\% with synthetic
queries and to 29.29\% with human queries. For Jina v4, the two give about the
same gains, and synthetic queries are slightly better on \DS{FiQA} for
$k\le 16$. On \DS{ViDoRe v3}, the matrix trained on 4,000 synthetic questions
from the target corpus improves baseline Ward by 1.1--5.3\%, against 0.1--1.0\%
for the matrix trained on 118,195 \DS{ColPali} queries from a different corpus.
Separately, the in-domain human-query matrix of \Cref{fig:multimodal-training}
gains 1.1--5.9\% on held-out queries (Ward, $k\le 32$). These comparisons use
different training data and evaluation query sets, so they do not isolate the
effect of synthetic versus human queries. The synthetic results use one seed.

\subsection{Reranking latency}
\label{sec:latency}

\begin{figure}[htb]
\centering
\includegraphics[width=0.5\linewidth,trim=0 12pt 0 4pt,clip]{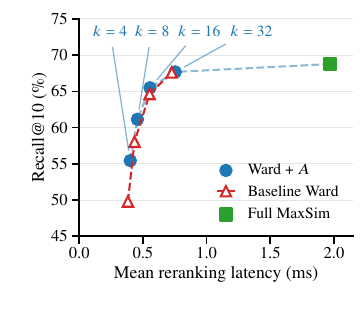}
\caption{ColBERTv2 Recall@10 versus mean single-core reranking latency on
  \DS{MS MARCO}, with $C=50$ candidates. Baseline Ward and Ward+$A$ use
  $k=4,8,16,32$ from left to right; full MaxSim is the uncompressed reference.
  Ward+$A$ points show means, with no error bars. Lines guide the eye.}
\label{fig:reranking-latency}
\end{figure}

\Cref{fig:reranking-latency} shows Recall@10 against reranking latency. Full
MaxSim reaches 68.77\% at 1.971\,ms per query. Ward at $k=4$ is about five
times faster but loses accuracy; the correction raises its Recall@10 from
49.79\% to 55.44\%. At $k=8$, it raises Recall@10 from 58.03\% to 61.14\%, at
4.3 times the speed of full MaxSim. Because $M$ is already applied to the
stored vectors, corrected and baseline Ward run the same operations per query.
At $k=32$, both are within 1.2 percentage
points of full MaxSim at about 2.6 times its speed.

\section{Conclusion}
We present a simple linear correction matrix to improve the performance of compressed representations in multi-vector embeddings. We show that this simple but effective method achieves consistent gains in text and visual document retrieval datasets, and can be implemented with synthetic queries as well. A systematic understanding of the mathematical procedure allowing this improvement would aid in the understanding of late-interaction models and help understand limits of such compression techniques.

\bibliographystyle{splncs04-unsrt}
\bibliography{references}

\appendix

\section{Ablations}
\label{app:ablations}
We report three ablations on \DS{MS MARCO} with ColBERTv2 and Ward
compression at $k\in\{4,8,16,32\}$. Each changes one part of the training
recipe in \Cref{sec:setup}: the number of training queries, the loss, or the
form of the correction. Unless stated otherwise, $A$ starts at the identity and
is trained with the KL loss of \Cref{eq:loss} ($T=1$, $\lambda=0.01$) and
Adam (learning rate $10^{-3}$, batches of 64 queries, no weight decay) for five
epochs, with 64-document panels built as in \Cref{sec:setup}. All runs use
training seed 0 only. We evaluate capped Recall@10 on all 6,980 dev queries
against the full corpus. For scale, Recall@10 SDs over five seeds are at most
0.20 points on \DS{MS MARCO} (\Cref{fig:MSMARCO-gains}).

\subsection{Number of training queries}
We train $A$ on nested subsets of 5,000, 50,000 and 100,000 \DS{MS MARCO}
training queries; the 5,000-query subset is the one used for the 
\DS{MS MARCO} results in the maintext. We evaluate the matrix after every epoch and do not
select an epoch. \Cref{tab:ablation-data} shows that more queries show diminishing returns.
At $k=4$, the epoch-5 matrix reaches 51.80\% with 5,000 queries, 52.35\% with
50,000 and 52.06\% with 100,000, against 45.53\% for baseline Ward. At larger
$k$, the three differ by less than 0.5 percentage points. Most of the gain is
already achieved within the first epoch.

\begin{table}[!htb]
\centering
\small

\begin{tabular}{@{}lrrrrr@{}}
\toprule
Training queries & Epoch & $k=4$ & $k=8$ & $k=16$ & $k=32$ \\
\midrule
Baseline & -- & 45.525 & 56.599 & 64.337 & 67.536 \\
\midrule
5,000 & 1 & 50.904 & 59.686 & 64.967 & 67.599 \\
 & 2 & 51.663 & 59.667 & 64.762 & 67.393 \\
 & 3 & 51.841 & 59.753 & 65.060 & 67.396 \\
 & 4 & 51.802 & 59.782 & 64.891 & 67.498 \\
 & 5 & 51.803 & 59.860 & 65.049 & 67.493 \\
\midrule
50,000 & 1 & 52.273 & 59.890 & 65.160 & 67.592 \\
 & 2 & 52.151 & 59.791 & 65.300 & 67.526 \\
 & 3 & 52.003 & 60.011 & 65.006 & 67.679 \\
 & 4 & 52.280 & 60.203 & 65.208 & 67.539 \\
 & 5 & 52.350 & 59.753 & 64.897 & 67.646 \\
\midrule
100,000 & 1 & 52.400 & 60.006 & 65.097 & 67.600 \\
 & 2 & 52.032 & 59.949 & 64.899 & 67.643 \\
 & 3 & 52.156 & 60.070 & 64.871 & 67.622 \\
 & 4 & 51.862 & 60.208 & 65.038 & 67.671 \\
 & 5 & 52.061 & 60.224 & 65.025 & 67.660 \\
\bottomrule
\end{tabular}
\caption{Number of training queries. Recall@10 (\%) of ColBERTv2 with Ward on
  \DS{MS MARCO} dev after each epoch; seed 0 and no epoch selection.
  Baseline is Ward without correction.}
\label{tab:ablation-data}
\end{table}

\subsection{Loss function}
With the 5,000-query subset, we compare 16 losses. Each also includes
$0.01\|A-I\|_F^2$, and we evaluate the matrix after epoch 5. Let $\hat p_Q$ be
the student's softmax distribution over the panel and $R_Q$ the judged
positives in the panel. The losses are:
\begin{itemize}
  \item \textbf{KL} (1 variant): the distillation loss of \Cref{eq:loss}.
  \item \textbf{Label} (1 variant): $-\log\sum_{D\in R_Q}\hat p_Q(D)$, which uses
    relevance judgments instead of teacher scores.
  \item \textbf{KL + label} (7 combinations): KL $+\,\beta\cdot$label, with fixed
    $\beta\in\{0.5,1\}$ or with $\beta\in\{0.25,0.5,1,2,4\}\cdot\beta_0$.
    Here $\beta_0$ is the ratio of the gradient norms of the KL and label
    losses with respect to $A$ at the identity, computed on the first 1,024
    training queries separately for each $k$.
  \item \textbf{Hinge + standardized KL} (7 combinations):
    $(1-\alpha)L_{\mathrm{hinge}}+\alpha\,\mathrm{KL}_z$ for
    $\alpha\in\{0,0.1,0.25,0.5,0.75,0.9,1\}$. Scores are standardized within
    each query's panel. The hinge applies a margin of 0.5 to the standardized
    student scores of 200 teacher-ordered document pairs per query, and
    $\mathrm{KL}_z$ is the KL loss between the standardized teacher and
    student scores.
\end{itemize}

\Cref{tab:ablation-loss} shows that the KL loss works best at $k=4$, reaching
51.80\%. Adding a small label term ($0.25\beta_0$) gives the same result
(51.79\%), and larger label weights lower Recall@10 steadily, to 50.28\% at
$4\beta_0$. The label loss alone gives 50.03\%, the hinge 49.14\% and
$\mathrm{KL}_z$ 50.55\%. The losses are closer at larger $k$; some combined
losses slightly exceed KL, and at $k=32$ all 16 are within 0.7 percentage
points.

\begin{table}[!htb]
\centering
\small
\begin{tabular}{@{}lrrrr@{}}
\toprule
Loss & $k=4$ & $k=8$ & $k=16$ & $k=32$ \\
\midrule
Baseline (no correction) & 45.525 & 56.599 & 64.337 & 67.536 \\
\midrule
KL & 51.803 & 59.860 & 65.049 & 67.493 \\
Label & 50.026 & 58.422 & 64.164 & 66.940 \\
KL $+\,0.25\beta_0\cdot$label & 51.794 & 59.996 & 65.094 & 67.400 \\
KL $+\,0.5\beta_0\cdot$label & 51.759 & 59.633 & 65.209 & 67.507 \\
KL $+\,\beta_0\cdot$label & 51.461 & 59.146 & 65.177 & 67.635 \\
KL $+\,2\beta_0\cdot$label & 50.683 & 58.756 & 64.902 & 67.525 \\
KL $+\,4\beta_0\cdot$label & 50.275 & 58.355 & 64.451 & 67.211 \\
KL $+\,0.5\cdot$label & 51.694 & 59.333 & 65.177 & 67.640 \\
KL $+\,1\cdot$label & 51.218 & 58.931 & 64.873 & 67.290 \\
\midrule
Hinge ($\alpha=0$) & 49.143 & 58.444 & 64.955 & 67.555 \\
Hinge $+\,\mathrm{KL}_z$ ($\alpha=0.1$) & 49.161 & 58.552 & 64.947 & 67.569 \\
Hinge $+\,\mathrm{KL}_z$ ($\alpha=0.25$) & 49.466 & 58.701 & 65.033 & 67.483 \\
Hinge $+\,\mathrm{KL}_z$ ($\alpha=0.5$) & 49.993 & 58.730 & 65.012 & 67.455 \\
Hinge $+\,\mathrm{KL}_z$ ($\alpha=0.75$) & 50.307 & 58.949 & 65.045 & 67.481 \\
Hinge $+\,\mathrm{KL}_z$ ($\alpha=0.9$) & 50.450 & 59.007 & 65.041 & 67.531 \\
$\mathrm{KL}_z$ ($\alpha=1$) & 50.546 & 59.128 & 65.119 & 67.416 \\
\bottomrule
\end{tabular}
\caption{Loss function. Recall@10 (\%) of ColBERTv2 with Ward on \DS{MS MARCO}
  dev, with 5,000 training queries, epoch 5 and seed 0. Every loss also
  includes $0.01\|A-I\|_F^2$.}
\label{tab:ablation-loss}

\end{table}

\subsection{Non-linear correction}
With the 100,000-query subset, we replace $A$ in \Cref{eq:score} with a
residual MLP $f(x)=x+g(x)$, applied to both query tokens and compressed
vectors. The map $g$ has 2, 3 or 4 linear layers with biases, hidden width 64,
128, 256 or 512 and GELU activations, and returns 128-dimensional vectors. Its
last layer starts at zero, so $f$, like $A$, starts at the identity. The MLPs
are trained with the KL loss and no regularization, and evaluated after epoch 5.
The $A$ reference is the regularized epoch-5 matrix trained on the same
100,000 queries (\Cref{tab:ablation-data}). The smallest MLP has about as many
parameters as $A$ (16,576 versus 16,384). Because $f$ is non-linear, it cannot
be moved to one side of the dot product like $M$: it must transform both the
query tokens and the stored vectors.

\Cref{tab:ablation-mlp} shows that every MLP scores above $A$ for $k\le16$.
At $k=4$, the smallest MLP reaches 54.00\%, against 52.06\% for $A$, and no
wider or deeper MLP exceeds 54.03\%. At $k=32$, all MLPs are within 0.25
percentage points of $A$. Because the MLPs are unregularized and $A$ is
regularized, this is not a comparison of architecture alone.

\begin{table}[!htb]
\centering
\small

\begin{tabular}{@{}lrrrrr@{}}
\toprule
Correction & Parameters & $k=4$ & $k=8$ & $k=16$ & $k=32$ \\
\midrule
Baseline (no correction) & 0 & 45.525 & 56.599 & 64.337 & 67.536 \\
$A$ & 16,384 & 52.061 & 60.224 & 65.025 & 67.660 \\
\midrule
MLP $2\times64$ & 16,576 & 54.003 & 60.893 & 65.883 & 67.690 \\
MLP $2\times128$ & 33,024 & 54.026 & 60.843 & 65.745 & 67.575 \\
MLP $2\times256$ & 65,920 & 53.754 & 60.649 & 65.591 & 67.536 \\
MLP $2\times512$ & 131,712 & 53.529 & 60.616 & 65.569 & 67.457 \\
\midrule
MLP $3\times64$ & 20,736 & 53.822 & 60.851 & 65.657 & 67.871 \\
MLP $3\times128$ & 49,536 & 53.916 & 60.652 & 65.802 & 67.702 \\
MLP $3\times256$ & 131,712 & 53.786 & 60.577 & 65.627 & 67.715 \\
MLP $3\times512$ & 394,368 & 53.580 & 60.556 & 65.503 & 67.615 \\
\midrule
MLP $4\times64$ & 24,896 & 53.415 & 60.714 & 65.456 & 67.818 \\
MLP $4\times128$ & 66,048 & 53.803 & 60.394 & 65.675 & 67.894 \\
MLP $4\times256$ & 197,504 & 53.702 & 60.955 & 65.392 & 67.753 \\
MLP $4\times512$ & 657,024 & 53.588 & 60.516 & 65.309 & 67.701 \\
\bottomrule
\end{tabular}
\caption{Non-linear correction. Recall@10 (\%) of ColBERTv2 with Ward on
  \DS{MS MARCO} dev, with 100,000 training queries, epoch 5 and seed 0. Depth
  counts the linear layers of $g$. MLPs are unregularized; $A$ includes
  $0.01\|A-I\|_F^2$. Parameters count only the correction.}
\label{tab:ablation-mlp}
\end{table}

\end{document}